\documentclass[10pt,a4paper,conference]{IEEEtran}
\usepackage{cite}

\ifCLASSINFOpdf
  \usepackage[pdftex]{graphicx}
  \graphicspath{{figures/}}
  \DeclareGraphicsExtensions{.pdf,.eps}
\else
\fi
\usepackage{amsmath}
\ifCLASSOPTIONcompsoc
 \usepackage[caption=false,font=normalsize,labelfont=sf,textfont=sf]{subfig}
\else
 \usepackage[caption=false,font=footnotesize]{subfig}

\usepackage{textcomp}
\usepackage{gensymb}
\usepackage{xurl}
\usepackage{multirow}
\usepackage[table,xcdraw]{xcolor}
\usepackage{soul}
\usepackage{booktabs}
\usepackage{balance}
\usepackage{array}
\usepackage{makecell}
\usepackage{fancyhdr}

\newif\ifinternalreview

\internalreviewfalse

\ifinternalreview
    \newcommand{\removed}[1]{\textcolor{red}{\textst{#1}}}
    \newcommand{\added}[1]{\textcolor{blue}{#1}}
    \newcommand{\Berriel}[1]{\textcolor{orange}{Berriel: #1}}
    
\else
    \newcommand{\removed}[1]{}
    \newcommand{\added}[1]{#1}
    \newcommand{\Berriel}[1]{}
\fi

\newcommand{\PreserveBackslash}[1]{\let\temp=\\#1\let\\=\temp}
\newcolumntype{C}[1]{>{\PreserveBackslash\centering}p{#1}}
\newcolumntype{R}[1]{>{\PreserveBackslash\raggedleft}p{#1}}
\newcolumntype{L}[1]{>{\PreserveBackslash\raggedright}p{#1}}
\newcommand{\mycell}[2]{\makecell{#1\\ \scriptsize{(#2 images)}}}

\begin{document}
%
\title{Deep Learning-based Type Identification of Volumetric MRI Sequences}

\author{
    \IEEEauthorblockN{
        Jean Pablo Vieira de Mello\IEEEauthorrefmark{1},
        Thiago M. Paixão\IEEEauthorrefmark{1}\IEEEauthorrefmark{2},
        Rodrigo Berriel\IEEEauthorrefmark{1},
        Mauricio Reyes\IEEEauthorrefmark{3},\\
        Claudine Badue\IEEEauthorrefmark{1},
        Alberto F. De Souza\IEEEauthorrefmark{1} and
        Thiago Oliveira-Santos\IEEEauthorrefmark{1}
    }
    \IEEEauthorblockA{
        \IEEEauthorrefmark{1}Universidade Federal do Espírito Santo (UFES), Brazil\\
        \IEEEauthorrefmark{2}Instituto Federal do Espírito Santo (IFES), Brazil\\
        \IEEEauthorrefmark{3}Artorg Center for Biomedical Engineering Research, University of Bern, Switzerland\\
        {\small \texttt{Email: jeanpvmello@gmail.com}}
    }
}


%


\maketitle

\begin{abstract}
The analysis of Magnetic Resonance Imaging (MRI) sequences enables clinical professionals to monitor the progression of a brain tumor. As the interest for automatizing brain volume MRI analysis increases, it becomes convenient to have each sequence well identified. However, the unstandardized naming of MRI sequences makes their identification difficult for automated systems, as well as makes it difficult for researches to generate or use datasets for machine learning research. In the face of that, we propose a system for identifying types of brain MRI sequences based on deep learning. By training a Convolutional Neural Network (CNN) based on 18-layer ResNet architecture, our system can classify a volumetric brain MRI as a FLAIR, T1, T1c or T2 sequence, or whether it does not belong to any of these classes. The network was evaluated on publicly available datasets comprising both, pre-processed (BraTS dataset) and non-pre-processed (TCGA-GBM dataset), image types with diverse acquisition protocols, requiring only a few slices of the volume for training. Our system can classify among sequence types with an accuracy of 96.81\%.
\end{abstract}


%
\IEEEpeerreviewmaketitle

\fancyhf{}
\renewcommand{\headrulewidth}{0pt}
\renewcommand{\footrulewidth}{0pt}
\fancyfoot[C]{\copyright 2021 IEEE.  Personal use of this material is permitted.  Permission from IEEE must be obtained for all other uses, in any current or future media, including reprinting/republishing this material for advertising or promotional purposes, creating new collective works, for resale or redistribution to servers or lists, or reuse of any copyrighted component of this work in other works.}
\thispagestyle{fancy}

\section{Introduction}

As the brain composes the center of the nervous system, associated diseases can affect skills, such as concentration, motor coordination, and memory as well as threaten life itself~\cite{amin2020brain}. Particularly, malignant brain tumors are among the most deadly cancer types, victimizing people of all ages when not early diagnosed and treated~\cite{sajjad2019multi}. The Magnetic Resonance Imaging (MRI) provides clinical experts with proper tridimensional visualization and analysis of the patient's brain tissues, allowing the detection of abnormalities~\cite{widmann2017mri,lenz2000methods,sajjad2019multi,tougaccar2020classification}. MR images are obtained by recording the intensities of the signals emitted by tissue's water protons when excited by a resonant electromagnetic radiofrequency field. Then, an image of the brain with visible contrast among its elements (e.g., fat, fluids, tumor) can be obtained by exploring their different characteristics (e.g., proton density and relaxation times)~\cite{widmann2017mri,lenz2000methods}. Usually, a better understanding of the abnormality is provided by the analysis of different types of MRI sequences, such as T1-weighted pre- and post-contrast-enhancement, T2-weighted and FLAIR~\cite{amin2020brain}.

\begin{figure}[t]
    \centering
    \includegraphics[width=\linewidth]{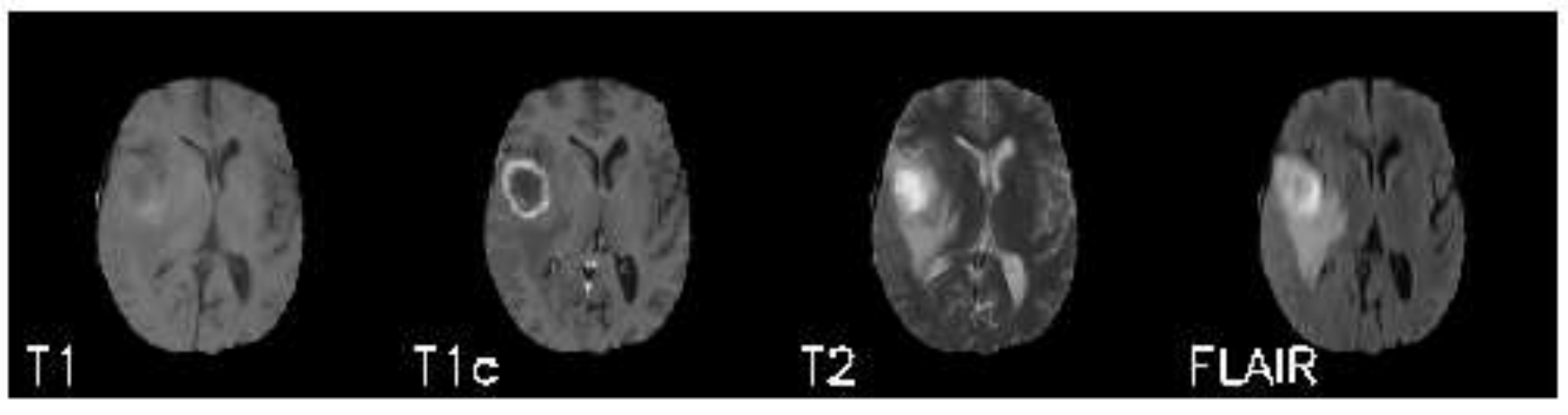}
    \caption{Examples of T1-weighted (T1), T1-weighted post-contrast-enhancement (T1c), T2-weighted (T2), and FLAIR MR images from the BraTS dataset referred to the same brain.} \label{fig1}
\end{figure}

Basically, T1-weighted images serve as material for analysis of the brain's healthy tissues when not contrast-enhanced by intravenous application of gadolinium in the analyzed patient. T1-weighted post-contrast-enhancement is similar to its counterpart, except that areas such as tumor borders are highlighted. In T2-weighted images, fluids present the highest intensities, including tumor edema. Finally, FLAIR images highlight the abnormality area while attenuating the cerebral fluid~\cite{currie2013understanding}. Figure~\ref{fig1} shows examples of MR images from The Multimodal Brain Tumor Image Segmentation Benchmark (BraTS) dataset contemplating these four sequence types referred to the same brain. 

The study of MRI sequences is not limited to the four previously presented types. Examples of other commonly used sequence types include the Diffusion-Weighted Images (DWI), for the study of the tissue microanatomy via water diffusion, Diffusion Tensor Images (DTI), a variant of DWI that takes into account the tissue's spatial volume, Perfusion-weighted images, for the adequate study of the veins, arteries, and vessels, and Proton Density-weighted (PD) images, for analysis of the water content in tissues \cite{widmann2017mri,tofts2003pd,alexander2007diffusion,lorio2019flexible}.

\begin{figure*}[t]
    \centering
    \includegraphics[width=\textwidth]{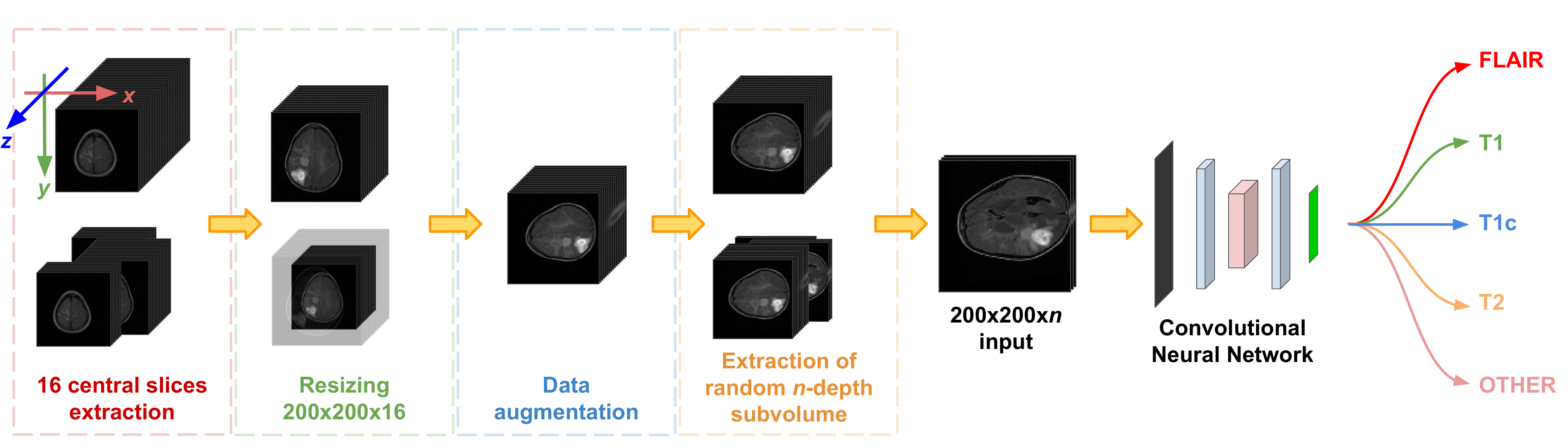}
    \caption{Overview of the proposed system. First, the 16 central slices of a volume are extracted. The 16-depth volume is resized to have slices with $200\times200$ pixels. The volume is augmented by the application of random rotation, translation, noise, brightness adjustment, and blurring. Then, a random subgroup of $n$ sequential slices ($n\leq16$) is picked from the 16-depth portion. This subgroup is provided as input to an 18-layer ResNet based CNN as an $n$-channel single image (illustration volume from the TCGA-GBM dataset).} \label{fig:overview}
\end{figure*}

As the interest for automatized processes for MRI processing increases alongside with the advances of machine learning, many works have been proposed involving automatization of tasks such as brain tumor segmentation and classification. Recently, Amin et al.~\cite{amin2020brain} proposed fusing T1-weighted, T1-weighted post-contrast-enhancement, T2-weighted and FLAIR sequences using discrete wavelet transform (DWT) so that the resulting image contains a more pronounced tumor area which is later segmented and classified as containing or not a tumor. Similarly, \"Ozyurt et al.~\cite{ozyurt2019brain} uses T1-weighted post-contrast-enhancement images to perform segmentation and classification of the tumor as benign or malign.

The dependency of the recent works on different sequence types faces an important issue: once obtained, the sequences must be properly annotated for their types. This is an expensive process in terms of working time and laborious efforts. Besides, the unstandardized annotations lead to an uncountable number of variations on the identification of the same category of data, which hinders the use of such data in machine learning. To tackle this issue, Noguchi et al.~\cite{noguchi2018artificial} and Ranjbar et al.~\cite{ranjbar2019deep} proposed methods for automatically annotating MR images by training a Convolutional Neural Network (CNN) for sequence type classification, treating the problem in a 2D perspective, avoiding the high memory and computational costs of 3D convolutions. However, both works have two limitations \added{that hinder comparability}: (i) classification is performed on individual slices and (ii) train/test is performed on private datasets assembled in partnership with clinical institutions \added{and (iii), sometimes, of limited size ~\cite{noguchi2018artificial}}. \added{The slice-based approach requires recurrent training/classification on slices that come from the same volume. Moreover, special care has to be taken when preparing the data of slice-based approaches to avoid unwanted bias in the results. For instance, when a dataset of individual slices is split into train, validation, and test sets, it should be explicitly guaranteed that slices from the same volume are in the same set; a guarantee that is not provided in~\cite{ranjbar2019deep}. Furthermore, this work presents missing and contradictory information relative to parameters such as the number of training epochs and the dataset size, which impairs reproducibility.}

To address the aforementioned issues, this work proposes an automated deep classifier that is capable of predicting the sequence type of a whole MRI volume input by using a 2D approach. The proposed system is trained and tested on publicly available datasets with high variability of data, which helps to confirm its robustness and enables the reproducibility of this research. Experimental results have shown that the system can distinguish sequence types of volumes provided by different centers and with different dimensions and orientations with an accuracy of 96.81\%.

The rest of the text is organized as follows. In Section \ref{sec:proposed}, the proposed system is presented. Section \ref{sec:experimental} details the experimental methodology. Following, the results are presented and discussed in Section \ref{sec:results}. Finally, Section \ref{sec:conclusion} concludes this work as well as discusses future perspectives.
\section{Proposed System}
\label{sec:proposed}

The proposed MRI sequence classifier uses \added{a Convolutional Neural Network (CNN) as the feature extractor. As CNNs are usually proposed to work with 3-channel images as inputs,} the first layer was modified to expect an $n$-channel image, where $n$ is a parameter specified by the user that denotes the volume depth, i.e., the number of adjacent slices in the MRI volume taken as input to the neural network. The output layer was also adjusted to decide among five classes, representing four of the most common sequence types that appear in the literature -- FLAIR, T1 (for T1-weighted images), T1c (for T1-weighted post-contrast-enhancement images), and T2 (for T2-weighted images) -- and an additional class labeled as OTHER (for sequences not belonging to any of the previous classes).

The proposed system was designed to work with MRI volumes of diverse formats and sizes. The following subsections describe how the data is loaded and transformed to provide simplified and standardized inputs to the CNN. An overview of the proposed system can be seen in Figure~\ref{fig:overview}.

\subsection{Data Loading}

The process of training/evaluation starts with the loading of each MRI volumes, whether they are stored in a single file or split into different files per slice. Each image is loaded once, normalized to fit the range $[0, 255]$, and has its 16 central slices extracted and stored into memory. If the volume contains less than 16 slices, each extreme slice is replicated in its respective direction until a 16-depth volume is created. Using only a central portion of the volume prevent some important issues:

\begin{itemize}
    \item the volume's extremities eventually contain slices with little representativeness for this work's purpose, while the central slices comprise a part of the brain volume with more relevant information;
    \item volumes stored as a set of separated slices have considerably slow loading, since each slice must be read and the volume with sorted slices must be assembled. Thus, an eager and unique loading of all data is way more efficient than loading each image whenever demanded;
    \item by considering only the volume's 16 central slices, the RAM memory overhead to store the data is reduced.
\end{itemize}

In order to reduce even more the memory allocation, all 16-depth extracted volumes have their height and width resized to 200 pixels each. Volumes with a height-width aspect ratio of 1:1 are resized normally. The volumes with a height different than width, however, are redimensioned to prevent information loss and deformation. Let $Min = min(height, width)$ and $Max = max(height, width)$. First, the volume is rescaled by a factor $R = Min/Max$ and then completed with zeros horizontally and vertically until both dimensions have $Min$ pixels. The resulting volume is finally resized to the resolution of $200\times200$ pixels and then added to the list of training/evaluation data.

\subsection{Data Transformation and Training}

The process of training the CNN is conducted by providing batches of randomly selected volumes from the just loaded dataset as input. Before each training epoch, a random subgroup of $n$ sequential slices is extracted from each 16-depth volume. This process aims at increasing data variability by picking different subgroups from the same volume in different epochs. To obtain even more data diversity, different augmentations are performed sequentially on each subgroup:
\begin{itemize}
    \item $Z$-axis rotation of $\alpha$ randomly drawn from $[-25\degree, 25\degree]$ (see Figure~\ref{fig:overview} for axes orientation);
    \item Another $z$-axis rotation of $i\times90\degree$, with $i$ randomly drawn from $\{0, 1, 2, 3\}$;
    \item Vertical translation of $\delta_y$ pixels randomly drawn from $[-\frac{\textrm{height}}{10}, +\frac{\textrm{height}}{10}]$, and  horizontal translation of $\delta_x$ pixels randomly drawn from $[-\frac{\textrm{width}}{10}, +\frac{\textrm{width}}{10}]$;
    \item Additive Gaussian Noise with zero mean and random standard deviation within $[0.0, 0.05\times255]$;
    \item Brightness adjustment by multiplying all pixels by a random factor within $[0.1, 2.0]$;
    \item Gaussian blurring with random sigma value within $[0.0, 1.0]$ with a probability of 50\%.
\end{itemize}

Before feeding the CNN, each slice from the final pixel data is standardized following a Gaussian distribution with zero mean and standard deviation of 1. At the end of each epoch, the model's macro-accuracy is evaluated on the validation set, and the final model is the one with the highest performance. The training parameters and protocols are presented in Section~\ref{subsec:expsetup}.

\subsection{\added{Inference}}
\added{At test time, unlike the random selection of a subgroup performed at each training step, the volume with the $n$ central slices is used. This central volume is then fed into the model, which predicts one of the five sequence types of interest: FLAIR, T1, T1c, T2, or OTHER. The predicted type is used to label the whole volume from which the subgroup came from.}
\section{Experimental Methodology}
\label{sec:experimental}

This section introduces the datasets used for training and evaluation of the model, the experimental setup, the conducted experiments, the evaluation metrics, and the computational resources used to conduct the experiments. Aiming at providing reproducible results, code and models are publicly available\footnote{\url{https://github.com/Jpvmello/type-identification-mri-sequences}}.

\subsection{Datasets}

Custom datasets based on datasets that are publicly available and commonly cited in the literature were assembled for the experiments. Section~\ref{sec:base} describes the base datasets used, while Section~\ref{sec:custom} describes the custom datasets assembled from them.

\subsubsection{Base Datasets}
\label{sec:base}

\begin{figure}[t]
    \centering
    \subfloat[]{\includegraphics[width=0.475\linewidth]{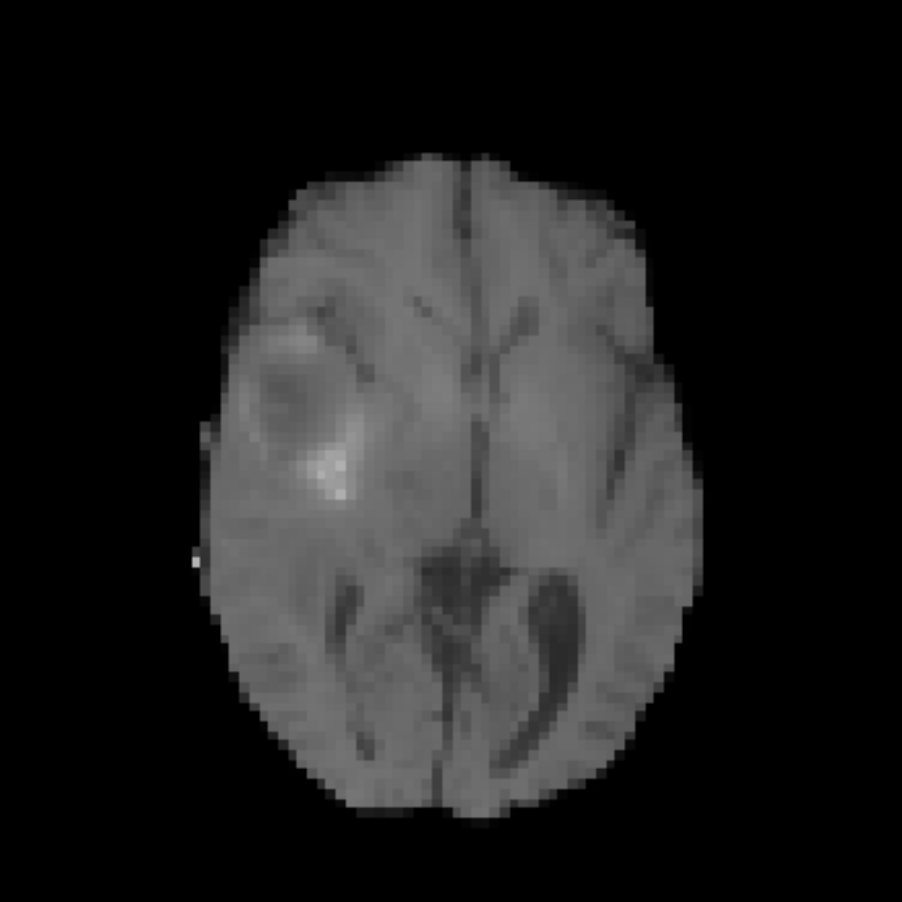}%
    \label{fig:brats}}
    \hfil
    \subfloat[]{\includegraphics[width=0.475\linewidth]{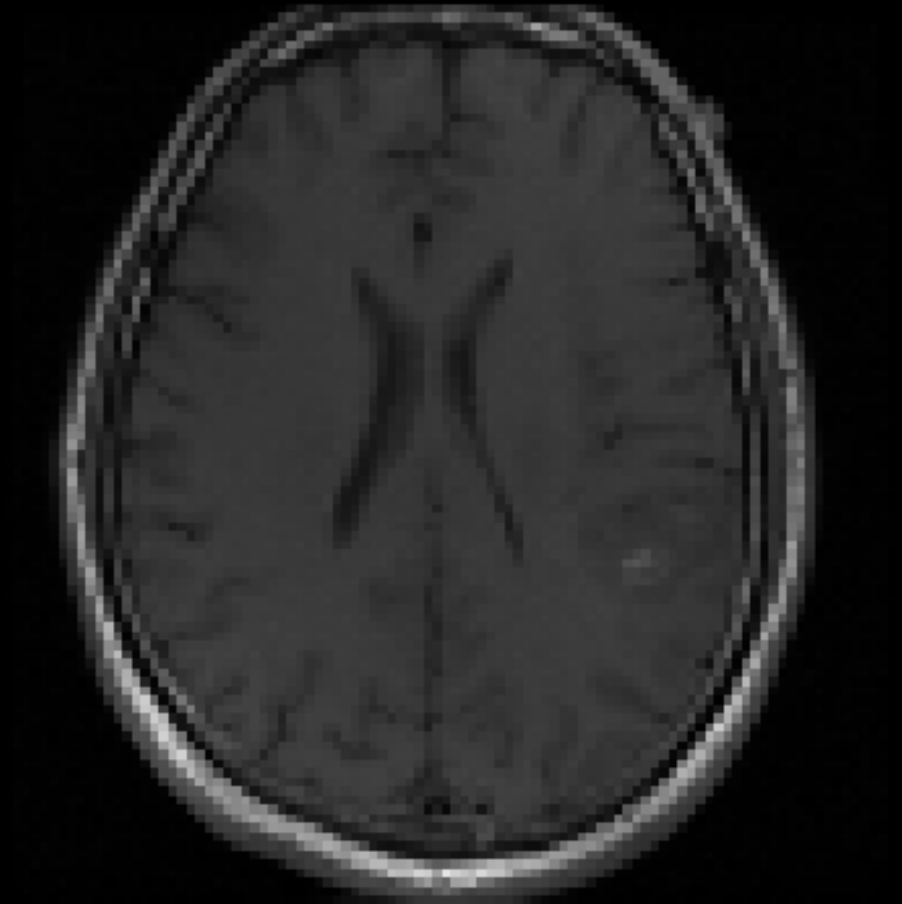}%
    \label{fig:tcga}}
    \caption{Example of (a) BraTS pre-processed image and (b) TCGA-GBM non-pre-processed data. Note that the visible area in pre-processed data corresponds only to the brain itself, while in non-pre-processed data the crane's skull is visible.}
    \label{fig:processing}
\end{figure}

\paragraph{The Multimodal Brain Tumor Image Segmentation Benchmark (BraTS)} Menze et al.~\cite{menze2014multimodal} proposed the BraTS dataset together with the organizers of the Medical Image Computing \& Computer-Assisted Intervention (MICCAI) 2012 and 2013 conferences, as a benchmark for the conference's tumor segmentation challenge, which takes place annually with eventual changes in BraTS data. The 2015 and 2019 versions of this dataset \cite{brats2019,brats2015,bakas2017advancing,bakas2018identifying,kistler2013virtual} were used for this work. Both datasets are publicly available and are originally split into training and validation sets with several study cases with one volume of each type among FLAIR, T1, T1c and T2. The training sets also contain one extra volume per case which is the ground truth segmentation of the tumor. A total of 1,840 volumes in NIfTI format equally distributed among the four MRI sequence types compose the BraTS 2019 dataset (1,340 from the training set and 500 from the validation set), while the 2015 version contains a total of 1,536 also balanced volumes in MetaImage format (1,096 from the training set and 440 from the validation set). The BraTS datasets provide pre-processed data, i.e., data previously subject to several processing steps. The most visible one is the skull-stripping, i.e., the removal of the crane's skull from the images. An example of BraTS pre-processed data is shown in Figure~\ref{fig:brats}.

\paragraph{The Cancer Genome Atlas Glioblastoma Multiforme (TCGA-GBM)} The Cancer Genome Atlas (TCGA) is a project which aims at reaching better comprehension, diagnosis, and prevention solutions to several types of cancer by cataloging their particular genomic causes and profiles and providing public datasets for the research community~\cite{tomczak2015cancer}. The TCGA-GBM dataset contains clinical, genomic, and MRI data from studies conducted with 262 patients from 8 different institutions. The imaging data consists of more than 480k DICOM slice files composing more than 5,000 volumes~\cite{klinger2019tcgagbm}. This dataset was used as a source of non-pre-processed volumes from several different scanning conditions and with high variability of sizes and orientations. Figure~\ref{fig:tcga} shows an example of a TCGA-GBM non-pre-processed image.

\subsubsection{Custom Datasets}
\label{sec:custom}

The custom datasets assembled, based on the BraTS and TCGA-GBM datasets, comprises five datasets: BRATS+TCGA5, BRATS+TCGA4, TCGA5, TCGA4, and BRATS4.

\paragraph{BRATS+TCGA5} BRATS+TCGA5 is considered the main dataset since it comprises images from all datasets and all considered classes. The ground truth labels were determined based on the name of the volume files, for the NIfTI and MetaImage data from BraTS, or of the base directory, for the TCGA-GBM DICOM data. BraTS file naming is well standardized and presented no challenge on interpretation, thus all brain volumes from the BraTS 2015 and 2019 training and validation sets were used. TCGA-GBM, however, presents considerable naming heterogeneity, which demanded careful extraction of information. Directories with terms ``FLAIR'' or ``T2'', regardless of letter case, were directly associated with those respective classes. Directories containing T1 and T1c data were differentiated based on terms such as ``PRE'', ``POST'', ``GD'' or ``GAD'' (referred to pre- or post-contrast-enhancement by infusion of gadolinium) and their combinations with the term ``T1''. Finally, data containing the terms ``DIFF'', ``DWI'', ``DTI'', ``PERF'' and ``PD'' were associated to the OTHER class, as the first three refer to diffusion sequence and the remaining ones to perfusion and proton-density sequences respectively. During the selection of the data, DICOM volumes with slices of mismatching dimensions were discarded. Table~\ref{tab:images_per_dataset} shows the number of volumes from each base dataset that were used to assemble BRATS+TCGA5. 10\% of the samples from each class and each base dataset were randomly selected to compose the BRATS+TCGA5 validation set, 20\% to the test set, and the remaining to the train set. The unbalanced samples from TCGA-GBM in the train set were oversampled such that all classes would have the number of samples as the class with originally most samples. Since the OTHER class does not occur in the BraTS sets, it was also oversampled to have the total number of samples of the remaining classes. Table~\ref{tab:main_sets} summarizes the validation, test, and train sets of the main dataset (summing BraTS sets for simplicity).

\begin{table}[ht]
    \centering
    \caption{Number of volumes from each base dataset that were used to assemble BRATS+TCGA5.}
    \label{tab:images_per_dataset}
    \resizebox{\columnwidth}{!}{
	    \begin{tabular}{@{}lcccccc@{}}
		    \toprule
		    Base dataset      & FLAIR & T1    & T1c   & T2    & OTHER & Total images \\ \midrule
		    BraTS'15 (train)  & 335   & 335   & 335   & 335   & 0     & 1,340        \\
		    BraTS'15 (val)    & 125   & 125   & 125   & 125   & 0     & 500          \\
		    BraTS'19 (train)  & 274   & 274   & 274   & 274   & 0     & 1,096        \\
		    BraTS'19 (val)    & 110   & 110   & 110   & 110   & 0     & 440          \\
		    TCGA-GBM          & 693   & 788   & 634   & 440   & 1,120 & 3,675        \\ \midrule
		    Total images      & 1,537 & 1,632 & 1,478 & 1,284 & 1,120 & 7,051        \\ \bottomrule
	    \end{tabular}
	}
\end{table}

\begin{table}[ht]
\centering
    \caption{Number of volumes from each base dataset that were used to assemble BRATS+TCGA5.}\label{tab:main_sets}
    \begin{tabular}{c|ccccc|c}
    \hline
    \multicolumn{7}{c}{BRATS+TCGA5 (main dataset): validation}                                             \\ \hline
    Base dataset & FLAIR        & T1           & T1c          & T2           & OTHER        & Total          \\ \hline
    \rowcolor[HTML]{EFEFEF}
    TCGA-GBM     & \textbf{69}  & \textbf{78}  & \textbf{63}  & \textbf{44}  & \textbf{112} & \textbf{366}   \\
    \rowcolor[HTML]{DFDFDF} 
    BraTS        & \textbf{83}  & \textbf{83}  & \textbf{83}  & \textbf{83}  & \textbf{0}   & \textbf{332}   \\
    \rowcolor[HTML]{CFCFCF} 
    Total        & \textbf{152} & \textbf{161} & \textbf{146} & \textbf{127} & \textbf{112} & \textbf{698}   \\ \hline
    \multicolumn{7}{c}{BRATS+TCGA5 (main dataset): test}                                                   \\ \hline
    Base dataset & FLAIR        & T1           & T1c          & T2           & OTHER        & Total          \\ \hline
    \rowcolor[HTML]{EFEFEF} 
    TCGA-GBM     & \textbf{139} & \textbf{159} & \textbf{128} & \textbf{88}  & \textbf{224} & \textbf{738}   \\
    \rowcolor[HTML]{DFDFDF}
    BraTS        & \textbf{172} & \textbf{172} & \textbf{172} & \textbf{172} & \textbf{0}   & \textbf{688}   \\
    \rowcolor[HTML]{CFCFCF} 
    Total        & \textbf{311} & \textbf{331} & \textbf{300} & \textbf{260} & \textbf{224} & \textbf{1,426} \\ \hline
    \multicolumn{7}{c}{BRATS+TCGA5 (main dataset): train}                                                  \\ \hline
                 & FLAIR        & T1           & T1c          & T2           & OTHER        & Total          \\ \hline
    \rowcolor[HTML]{EFEFEF} 
    TCGA-GBM     & \textbf{}    & \textbf{}    & \textbf{}    & \textbf{}    & \textbf{}    & \textbf{}      \\
    \rowcolor[HTML]{EFEFEF} 
    Original     & 485         & 551          & 443          & 308          & 784          & 2,571         \\
    \rowcolor[HTML]{EFEFEF} 
    Oversampled  & 299         & 233          & 341          & 476          & 589          & 1,938         \\
    \rowcolor[HTML]{EFEFEF} 
    Total & \textbf{784}   & \textbf{784}   & \textbf{784}   & \textbf{784}   & \textbf{1,373}  & \textbf{4,509} \\
    \rowcolor[HTML]{DFDFDF} 
    BraTS        & \textbf{589} & \textbf{589} & \textbf{589} & \textbf{589} & \textbf{0}   & \textbf{2,356} \\
    \rowcolor[HTML]{CFCFCF} 
    Total & \textbf{1,373} & \textbf{1,373} & \textbf{1,373} & \textbf{1,373} & \textbf{1,373} & \textbf{6,865} \\ \hline
    \end{tabular}
\end{table}

\paragraph{BRATS+TCGA4} The variant BRATS+TCGA4 contains all images from the main dataset, except those which belong to the OTHER class, i.e., all images from the four mainly studied classes FLAIR, T1, T1c, and T2. The validation set is reduced from 698 (see Table~\ref{tab:main_sets}) to 586 images, the test set from 1,426 to 1,202 images, and the train set from 6,865 to 5,492 images uniformly balanced among the four classes.

\paragraph{TCGA5} It contains the TCGA-GBM data from the main dataset and, therefore, possesses only non-preprocessed data. The validation and test sets contain 366 and 738 images, respectively. For the train set, the oversampling of the OTHER class was neglected, so that all classes have 784 samples each (3,920 images, in total).

\paragraph{TCGA4} Similar to TCGA5, but excluding the samples from the OTHER class. The validation, test, and training sets contain 254, 514, and 3,136 images each.

\paragraph{BRATS4} BRATS4 contains data from all BraTS sets, naturally contemplating only the four classes FLAIR, T1, T1c, and T2. This dataset contains exclusively pre-processed data. The validation, test, and train sets are composed of 332, 688, and 2,356 images each.

Finally, it is worth mentioning that there is no intersection between validation, test, and train sets both within a single dataset and among all datasets.

\subsection{Experimental Setup}
\label{subsec:expsetup}

The training during the experiments was conducted according to the following parametrization: the batch size of 32, a learning rate of 0.01, and 300 training epochs. Also, it was adopted the cross-entropy function as loss criterion and the Stochastic Gradient Descent (SGD) with a momentum of 0.9 as the optimization algorithm.

\subsection{Evaluation Metrics}

The performance of the model was evaluated based on its validation and testing macro-accuracies, as well as on the testing confusion matrix.

\subsection{Experiments}

The experiments aim at finding the model which presents the best generalization when predicting MRI sequence types. To achieve this, studies were conducted on (i) the optimal value of $n$, i.e., the depth of the input volume; and (ii) the influence of using pre-processed training data. In all cases, the best-trained model is obtained from the training epoch with the highest validation accuracy among the 300 training.

\subsubsection{Study on the Input Volume Depth}

For the first study, the main dataset BRATS+TCGA5 was used for training, validation and testing for $n = 1, 2, 3, \ldots, 16$. The validation accuracy was computed for each of these experiments. \added{Note that, because the $n$ central slices are used at inference time, when $n$ is even, the volume is perfectly centralized, while that is not the case when $n$ is odd. Therefore, the validation accuracy for odd values of $n$ is given by the average of both volumes closest to the center of the original volume. The goal of this study is to investigate whether or not the depth of the input volume has an impact on the results.} The value of $n$ that yielded the highest validation accuracy was then adopted for the second study. \added{Moreover, given that the proposed method is not limited to a specific CNN, in addition to the 18-layer ResNet~\cite{he2016deep} used in the input volume depth study, a comparison with other well-known CNN architectures (AlexNet~\cite{alexnet}, SqueezeNet 1.1~\cite{squeezenet}, MobileNetV2~\cite{mobilenetv2}, and VGG16~\cite{vgg}) is also performed. These models followed a similar experimental setup (for some of them the learning rate was reduced to 10\% of the value used for ResNet-18, aiming at convergence).}

\subsubsection{Study on the Use of Pre-Processed Data}

As an analysis of the influence of pre-processed data to the performance, the datasets with pre-processed data (BRATS4), non-pre-processed data (TCGA5 and TCGA4), and both kinds of data (BRATS+TCGA5 and BRATS+TCGA4) were trained (validated by their corresponding validation set) and tested mutually. The experiment respected the number of classes comprised by each dataset, i.e., the training on each four-class dataset (BRATS4, TCGA4, and BRATS+TCGA4) was tested exclusively and individually on the four-class datasets, while the training on each five-class dataset (TCGA5 and BRATS+TCGA5) was tested on each five-class dataset only.

\subsection{Computational Resources}

Both training and evaluation processes were performed on an Intel\textsuperscript{\textregistered} Core\textsuperscript{\texttrademark} i7-4770 CPU (3.40GHz, 16GB RAM), an Intel\textsuperscript{\textregistered} Xeon\textsuperscript{\textregistered} CPU E5606 (2.13GHz, 22GB RAM) and an Intel\textsuperscript{\textregistered} Xeon\textsuperscript{\textregistered} CPU X5690 (3.47GHz, 50GB RAM). The first two are supported by an NVIDIA TITAN Xp GPU and the last one by an NVIDIA GeForce TITAN X GPU, all with 12GB memory. 
\section{Results and Discussion}
\label{sec:results}

\subsection{Study on the Input Volume Depth}

\begin{figure}[h]
    \centering
    \includegraphics[width=0.48\textwidth]{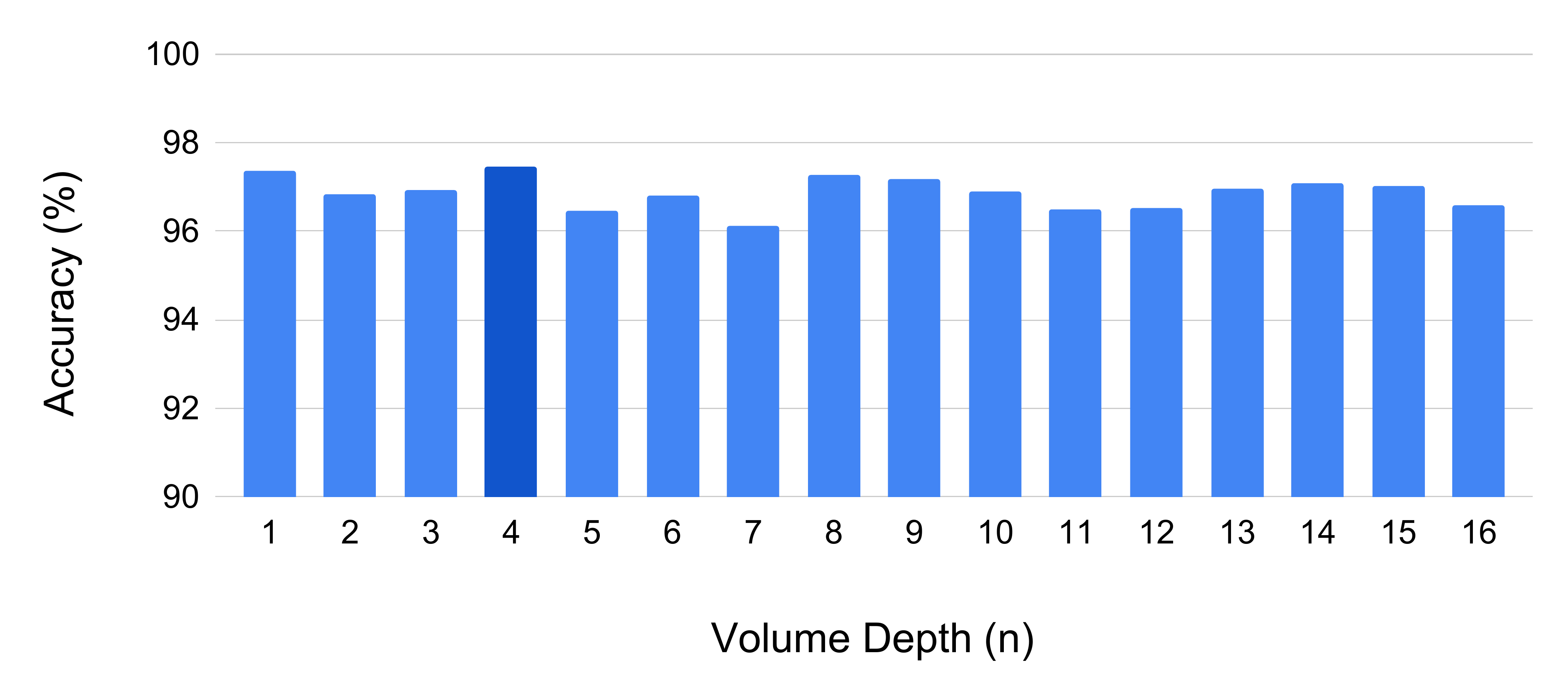}
    \caption{\added{Validation accuracy for the model trained on BRATS+TCGA5 considering several different values of $n$ (input volume depth). The highest performance is highlighted in dark blue ($n=4$).}}
    \label{fig:accxn}
\end{figure}

The validation accuracy, for the model trained and evaluated on BRATS+TCGA5, as a function of $n$ is shown in Figure~\ref{fig:accxn}.
\added{As can be seen, the validation performance does not seem to follow a specific trend, i.e., deeper input volumes sometimes improve the performance, but are eventually worse than using shallow ones.}
This behavior is within expectations, given that, on one hand, smaller values of $n$ result in more combinations of non-repeated slices at training time, increasing training data variability; and, on the other hand, higher $n$ means more information for the model to learn.

It is worth mentioning that the difference between the lowest and highest accuracies is very small: 1.32 p.p. for validation (96.12\% at $n = 7$ and 97.44\% at $n = 4$).
Moreover, the results show that few slices seem to be enough to characterize a volume, and using more slices does not seem to considerably impair the sequence identification. \added{For the next experiments, we chose the model with $n = 4$, since it achieved the highest validation accuracy, achieving a test accuracy of $96.81\%$.}

\added{To further investigate the robustness of the proposed approach, several other well-known CNN architectures were investigated, besides the ResNet-18 used in the aforementioned study. This is only possible because our approach is not limited to a specific architecture. In this comparison, the models were trained and evaluated with the best $n$ found in the previous experiment: $n=4$. The results, in terms of macro-accuracy, are shown in Table~\ref{tab:net_comparison}, where all of them presented high and comparable accuracy. Among the investigated architectures, it still seems reasonable to keep using ResNet-18 for the next experiments, because it is, in terms of the number of parameters, not as simple as SqueezeNet and MobileNet nor as complex as AlexNet or VGG16,  reducing the probability of underfitting/overfitting~\cite{lever2016points}.}

\begin{table}[h]
    \centering
    \caption{\added{Comparison between the ResNet-18's validation accuracy and other four well-known CNN architectures.}}
    \label{tab:net_comparison}
    \begin{tabular}{@{}lc@{}}
    \toprule
    Architecture    & Accuracy (\%) \\ \midrule
    SquezeeNet 1.1  & 96.02         \\
    AlexNet         & 96.29         \\
    MobileNetV2     & 96.88         \\
    VGG16           & 97.61         \\
    \midrule
    ResNet-18       & 97.44         \\
    \bottomrule
    \end{tabular}
\end{table}

\subsection{Study on the Use of Pre-Processed Data}

For this study, $n$ was set to 4, which was the value that yielded the highest validation accuracy on the main dataset (97.44\%) in the previous study. The test accuracy obtained for each possible combination of train-test datasets, considering only the four-class datasets, is shown in Table~\ref{tab:processing4}. The results provide evidence that a model trained individually with each pre-processed data (BRATS4) or non-pre-processed data (TCGA4) performs worse as the proportion of the same kind of data in the test decreases: training with BRATS4 yields 99.27\% of accuracy when testing with the same kind of data (BRATS4), 82.90\% when testing with mixed data (BRATS+TCGA4) and 62.44\% when tested on data of the opposite kind (TCGA4), while training with TCGA4 yields 92.99\% for tests with the same kind of data, 65.01\% for mixed data and 42.59\% for the opposite type of data.

\begin{table}[h]
    \centering
    \caption{Testing accuracy (\%) for each possible combination of four-class train-test datasets with $n = 4$.}\label{tab:processing4}
    \begin{tabular}{c|ccc}
    \toprule
                   & \multicolumn{3}{c}{Test Datasets} \\ \cmidrule{2-4}
    Train Datasets & \mycell{BRATS4}{688} & \mycell{BRATS+TCGA4}{1,202} & \mycell{TCGA4}{514} \\ \midrule
    \mycell{BRATS4}{2,356}	     & 99.27 & 82.90 & 62.44 \\
    \mycell{BRATS+TCGA4}{5,492}  & 98.98 & 96.36 & 93.30 \\
    \mycell{TCGA4}{3,136}	 	 & 42.59 & 65.01 & 92.99 \\ \bottomrule
    \end{tabular}
\end{table}

This effect is also seen in Table~\ref{tab:processing5}, which shows the test accuracy for the combinations of the train-test of the five-class datasets. A model trained on the TCGA5 performs well (94.66\% accuracy) when tested on the same kind of data, while testing with mixed data reduces the accuracy to 52.68\%.

\begin{table}[h]
    \centering
    \caption{Testing accuracy (\%) for each possible combination of five-class train-test datasets with $n = 4$.}\label{tab:processing5}
    \begin{tabular}{c|cc}
    \toprule
    		                    & \multicolumn{2}{c}{Test Datasets} \\ \cmidrule{2-3} 
    Train Datasets 				& \mycell{BRATS+TCGA5}{1,426} & \mycell{TCGA5}{738} \\ \midrule
    \mycell{BRATS+TCGA5}{6,865} & 96.81 					  & 94.51 \\
    \mycell{TCGA5}{3,920}       & 52.68 					  & 94.66 \\ \bottomrule
    \end{tabular}
\end{table}

\begin{figure}[h]
    \centering
    \includegraphics[width=\columnwidth]{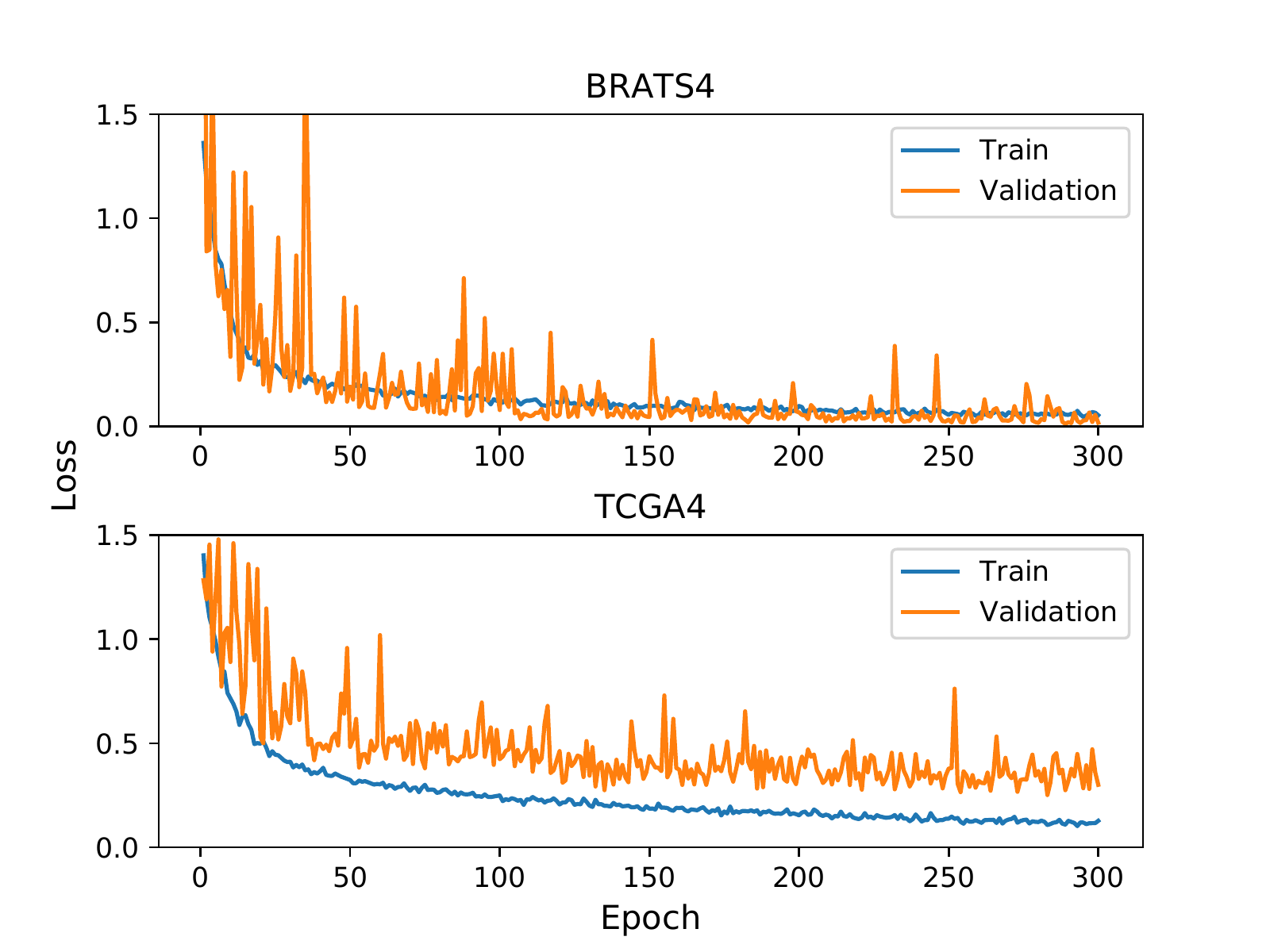}
    \caption{\added{Training and validation loss for training with BRATS4 (pre-processed data, top) and with TCGA4 (non-pre-processed data, bottom).}}
    \label{fig:losses}
\end{figure}

Referring to the models trained with mixed data (BRATS+TCGA4 and BRATS+TCGA5), the model trained on the four-class dataset (BRATS+TCGA4) presents 98.98\% of accuracy on pre-processed data only (BRATS4), 96.36\% on mixed data (BRATS+TCGA4) and 93.30\% on non-pre-processed data only (TCGA4), while its five-class correspondent (trained on BRATS+TCGA5) yields 96.81\% testing accuracy on mixed data (BRATS+TCGA5) and 94.51\% on non-pre-processed data (TCGA5). These results point that there is a decrease in accuracy which follows the proportion of non-processed-data. However, this is a dataset issue. In fact, BraTS data is standardized in terms of volume original dimensions and orientation, which makes its data easier to learn and recognize, while TCGA-GBM is affected by its significant data heterogeneity. This also explains why, among the models trained with only one kind of data, the BraTS-based one presents better generalization. \added{This becomes even more visible when looking at the behavior of the training and validation losses for pre-processed data only (as BRATS4) compared to the ones for non-pre-processed data only (as TCGA4), as shown in Figure~\ref{fig:losses}. It is noticeable that the validation loss behaves nearly the same as the training loss for the pre-processed data, while it presents a significant slower decrease for the non-pre-processed data.}

Overall, the results show that the models trained with mixed data presented the best performance on most test datasets. The only exceptions were the models trained and tested using BRATS4 and the one trained and tested using TCGA5. Nevertheless, the results are extremely close, with less than 0.5 p.p. difference in both cases when compared with their corresponding training with mixed data.
Also, the models trained with mixed data maintain high accuracy levels within a minimum higher than 93\% (BRATS+TCGA4 tested on TCGA4) and a maximum close to 99\% (BRATS+TCGA4 tested on BRATS4).


\subsection{Discussion on the System Performance}

\begin{figure*}[t]
    \centering
    \includegraphics[width=\textwidth]{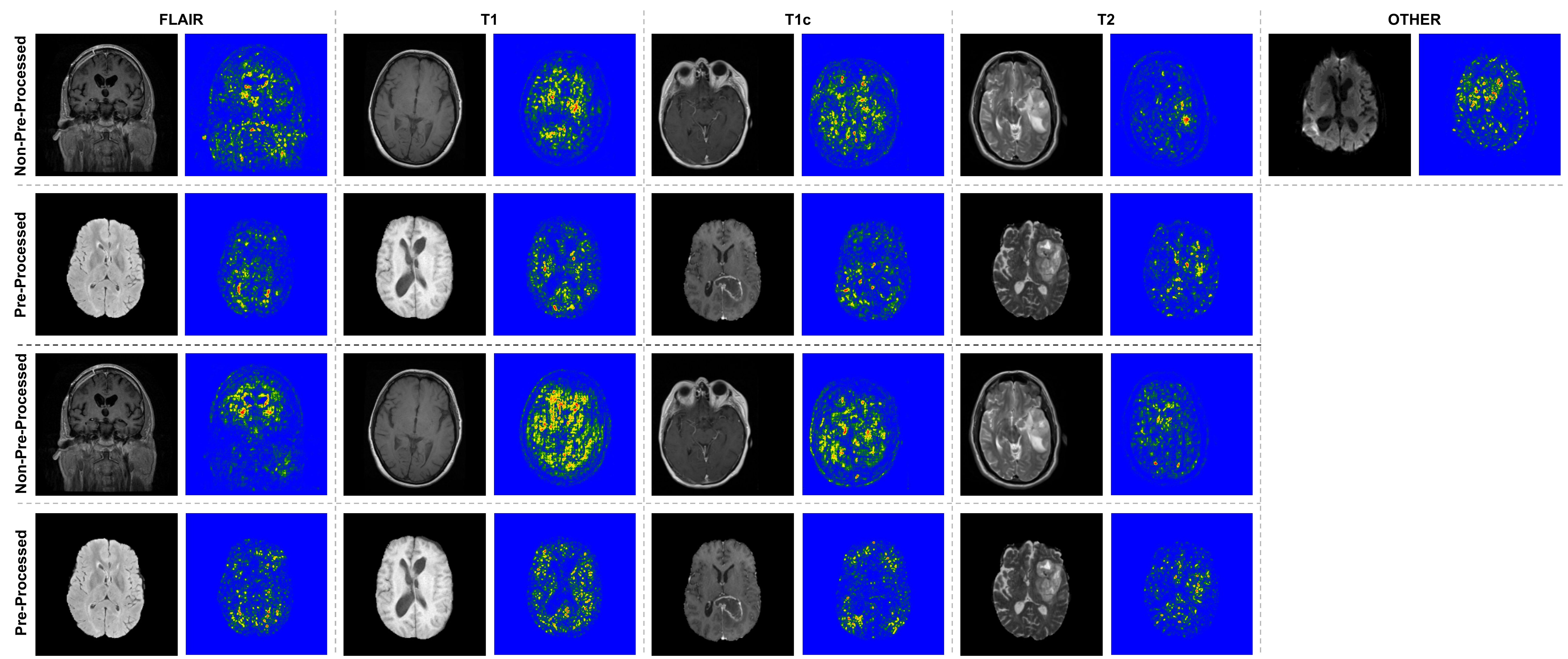}
    \caption{\added{Relevant features (based on Integrated Gradients) of volumes from all considered classes and pre-processing conditions. The first two rows come from the model trained on mixed data (BRATS+TCGA5), while the two last rows come from models trained on TCGA4 (3rd row) and BRATS4 (4th row).}}
    \label{fig:maps}
\end{figure*}

The second study revealed that the models trained with both pre-processed and non-pre-processed data, BRATS+TCGA4 and BRATS+TCGA5 (main dataset), achieved better generalization than models trained with only one kind of data. As the main dataset (including the OTHER class) is more complete and complex than its four-class correspondent, the best model was considered to be the BRATS+TCGA5 trained with the main dataset with $n = 4$.


Overall, the confusion matrices show that the main factor that negatively impacts accuracy is the difficulty in distinguishing T1 and T1c images, as the latter type is a direct variation of the former one. For the test involving the main dataset with $n = 4$, 26/331 T1 images were misclassified as T1c and 4/300 T1c images were predicted as T1, as shown in Table~\ref{tab:matrix_main_main}. Knowing this issue provides a direction for future improvements to the system's accuracy.

\begin{table}[!ht]
    \centering
    \caption{Confusion matrix for training and testing with the main dataset with $n = 4$.}
    \label{tab:matrix_main_main}
    \renewcommand{\arraystretch}{2}
    \addtolength{\tabcolsep}{-2pt}
    \begin{tabular}{ll|C{0.9cm}|C{0.9cm}|C{0.9cm}|C{0.9cm}|C{0.9cm}|}
        \multicolumn{2}{c}{}&   \multicolumn{5}{c}{Predicted Classes}\\
        \multicolumn{2}{c}{}& \multicolumn{1}{c}{FLAIR} & \multicolumn{1}{c}{T1} & \multicolumn{1}{c}{T1c} & \multicolumn{1}{c}{T2} & \multicolumn{1}{c}{OTHER}\\
        \cline{3-7}
        \multirow{5}{*}{{\rotatebox[origin=c]{90}{Actual Classes}
        }} & FLAIR & 307  & 2    & 2    & 0    & 2  \\ \cline{3-7}
        &   T1     & 1    & 303  & 26   & 1    & 0 \\ \cline{3-7}
        &   T1c    & 1    & 4    & 293  & 1    & 1 \\ \cline{3-7}
        &   T2     & 0    & 0    & 0    & 258  & 2 \\ \cline{3-7}
        &   OTHER  & 1    & 2    & 2    & 2    & 217 \\ \cline{3-7}
    \end{tabular}
    \addtolength{\tabcolsep}{2pt}
\end{table}

 \added{\textbf{Qualitative analysis of the features.} In addition to the quantitative experiments, a qualitative analysis was also performed. The goal of this experiment is to better understand what was learned by the models, especially for the different pre-processing conditions. For this analysis, models (all with $n=4$) trained on three different datasets were used: BRATS+TCGA5, TCGA4, and BRATS4. The features were extracted from samples that belong to the test set of the respective dataset used for the training. For extracting the relevant features, a well-known explainable AI technique called Integrated Gradients~\cite{ig2017icml} (IG) was applied. The results based on IG for one sample of each sequence type and each pre-processing condition are shown in Figure~\ref{fig:maps}. Although it may not be clear on what characterizes each sequence type, it is noticeable that the relevant features are concentrated on the tissues rather than the tumor, edema, fluids, skull, eyes, or any other elements represented within the volumes. This may also be the reason for the correct predictions on slices that do not present tumor, such as the non-pre-processed FLAIR sample shown in Figure~\ref{fig:maps}. Besides, these findings may provide an insight into the accuracy differences between the different datasets (pre-processed and non-pre-processed data, as well as mixed and non-mixed data). For example, the non-pre-processed data presents many more elements apart from the relevant tissues, such as skull and eyes, that are not present on the pre-processed data.}

\section{Conclusion}
\label{sec:conclusion}

The analysis of different types of MRI sequences is essential for the study and diagnosis of the multiple variations of brain abnormalities. However, the manual work of documenting large numbers of MRI sequences is laborious and unstandardized, making difficult the data collecting for automatic algorithms for image visualization and analysis. This encourages the proposal of an automated MRI sequence type classifier.

The experimental results show that the training of an 18-layer ResNet-based CNN enables the classification of MRI volumes with different formats, dimensions, orientations, and acquisition processes among FLAIR, T1, T1c, T2, and other sequence types with an accuracy of up to 96.81\%. This can be achieved by training with small amounts of slices from potentially deep volumes. It was also verified that the classifier presents good performance when trained and tested with only pre-processed data 
or with only non-pre-processed data
, while it does not perform very well when tested on data with sequence types different from those used for training. These results indicate that using heterogeneous data as in BRATS+TCGA5 can provide a good trade-off, which is further confirmed by the experiments with mixed data training.

A further improvement involves investigating procedures to distinguish even better between very similar sequences such as T1 and T1c. Increasing the number of different sequence types distinguishable by the classifier is also feasible, supporting clinical studies with more content and minimal manual efforts.
\section*{Acknowledgements}
\label{sec:acknowledgements}

We gratefully acknowledge the funding received from the Swiss Cancer League (grant KFS-3979-08-2016), Coordenacão de Aperfeiçoamento de Pessoal de Nível Superior - Brasil (CAPES) - Finance Code 001, Conselho Nacional de Desenvolvimento Científico e Tecnológico (CNPq, Brazil) and Fundação de Amparo à Pesquisa do Espírito Santo - Brasil (FAPES) (grant 84412844). This project was supported by the Swiss Personalized Health Network (SPHN) initiative. Also, the authors thank NVIDIA Corporation for the donation of the GPUs used for this work.

\balance

\bibliographystyle{IEEEtran}
\bibliography{paper}

%



\newcommand{\todo}[1]{\textcolor{red}{#1}}
\newcommand{\question}[1]{\vspace{5pt} \noindent \textbf{#1:}}
\newcommand{\answer}[1]{\noindent #1}
\end{document}

